\documentclass[11pt]{article}
\usepackage[a4paper,left=1.5cm,right=1.5cm,top=2cm,bottom=2cm]{geometry}
\usepackage[utf8]{inputenc}
\usepackage[T1]{fontenc}
\usepackage{lmodern}

\usepackage{amsmath, amsthm, amscd, amsfonts, amssymb}
\usepackage{graphicx}
\usepackage{float}
\usepackage{subcaption}
\usepackage{multirow}
\usepackage{url}
\usepackage{xcolor}

\usepackage{hyperref}
\hypersetup{
	colorlinks=true,
	linkcolor=blue,
	citecolor=blue,
	urlcolor=blue
}

\usepackage{pgfplots}
\usepackage{pgfplotstable}
\usepackage{lmodern}
\usepackage{tikz}                     \usepackage{tikzsymbols}
\usepackage{tikz-cd}
\usetikzlibrary{shapes.geometric, arrows.meta, automata,positioning, calc}

\newtheorem{theorem}{Theorem}[section]

\theoremstyle{definition}

\theoremstyle{remark}
\newtheorem{remark}[theorem]{Remark}
\numberwithin{equation}{section}

\title{A chaotic flux cipher based on the random cubic family $f_{c_n}(z)=z^3+c_n z$}

\author{
	P. Mehdipour\\
	Universidade Federal de Viçosa, Brazil\\
	\texttt{pouya@ufv.br}
	\and
	A. Miranda Alves\\
	Universidade Federal de Viçosa, Brazil\\
	\texttt{amalves@ufv.br}
	\and
	G. Honorato\\
	Universidad de Valparaíso, Chile\\
	\texttt{gerardo.honorato@uv.cl}
	\and
	M. Salarinoghabi\\
	Universidade Federal de Viçosa, Brazil\\
	\texttt{mostafa.salarinoghabi@ufv.br}
}

\date{} 

\begin{document}
	
	\maketitle

\begin{abstract}
This paper presents a symmetric stream cipher that utilizes the dynamic properties of random cubic mappings in the complex plane to generate pseudo-random key streams. The system is based on the iterations of the random cubic polynomial $f_n(z)=z^3+c_n z$, where the parameters $c_n$ are chosen randomly from a disc of radius $\delta$ and with center at the origin, aiming to improve the chaotic behaviour and, consequently, the randomness of the generated sequence. The stability of the Julia set under small parameter perturbations, when $\delta < \delta_0\simeq 0.89$, is considered to ensure key consistency in noisy environments, such as 5G networks. On the other hand, for $\delta > 3$, the system exhibits instability and chaos, ideal for generating ultra-secure keys. The Python implementation integrates secure key derivation, robust key stream generation via warmed-up iteration, and an authenticated encryption scheme using the modern cryptographic primitives (\texttt{HKDF} and\texttt{HMAC-SHA-256}), to ensure message integrity and authenticity. Statistical analyses, including chi-square test and entropy calculation, are performed on the output of the key stream generator to evaluate its randomness and distribution. 
In addition, a complete statistical validation, compliant with \texttt{NIST SP 800-22} standards in modern cryptography, was performed to enhance the proposed system's credibility.
\end{abstract}


\section{Introduction}\label{sec:intro}
The security of digital communications is a basis of modern technological infrastructure. Symmetric cryptography, in particular stream ciphers, plays an important role in providing confidentiality with high efficiency. A stream cipher is a symmetric encryption technique where data is encrypted one bit or byte at a time, rather than in blocks. It uses a Pseudo-random keystream, generated from a secret key, to combine with the plaintext.

The effectiveness of a stream cipher critically depends on the quality of its Pseudo-Random Number Generator (\texttt{PRNG}), which must produce unpredictable, non-repetitive, and statistically random sequences. Traditionally, cryptographic \texttt{PRNG}s are built on computationally difficult mathematical problems. However, the exploration of chaotic dynamical systems, in particular random dynamical systems,  has became as a promising area for the development of new cryptographic primitives, due to their sensitivity to initial conditions and complex and unpredictable behaviour.

This work investigates on the application of random iterative mappings in the complex plane for the generation of cryptographic key streams. Specifically, we focus on the family of random cubic polynomials defined by $f_{c_n}(z) = z^3 + c_nz$, where the parameters $c_n$ are chosen from a disk of radius $\delta$ and they vary randomly at each iteration. To be more precise, if we denote the sequence $(c_n)$ by $\omega$ then the $m^{\text{th}}$-iteration of a point $z_0\in\mathbb{C}$ under the map $f_{c_n}$ is given by
\begin{eqnarray}\label{eq:Fn}
f^m_{\omega}(z_0) = f_{c_m}\circ\dots\circ f_{c_1}(z_0).
\end{eqnarray}
This dynamic variation of the parameters aims to intensify the chaotic behaviour of the system, resulting in sequences with improved randomness properties. Recent literature has demonstrated the potential of chaotic systems for cryptographic applications \cite{alvarez2020some, sun2020chaotic}.

\begin{figure}[h!]
	\centering
	\includegraphics[width=0.23\linewidth]{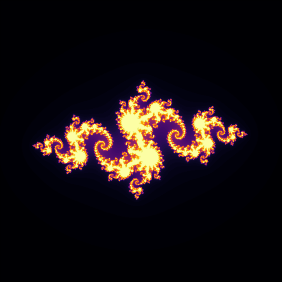}
	\qquad
	\includegraphics[width=0.23\linewidth]{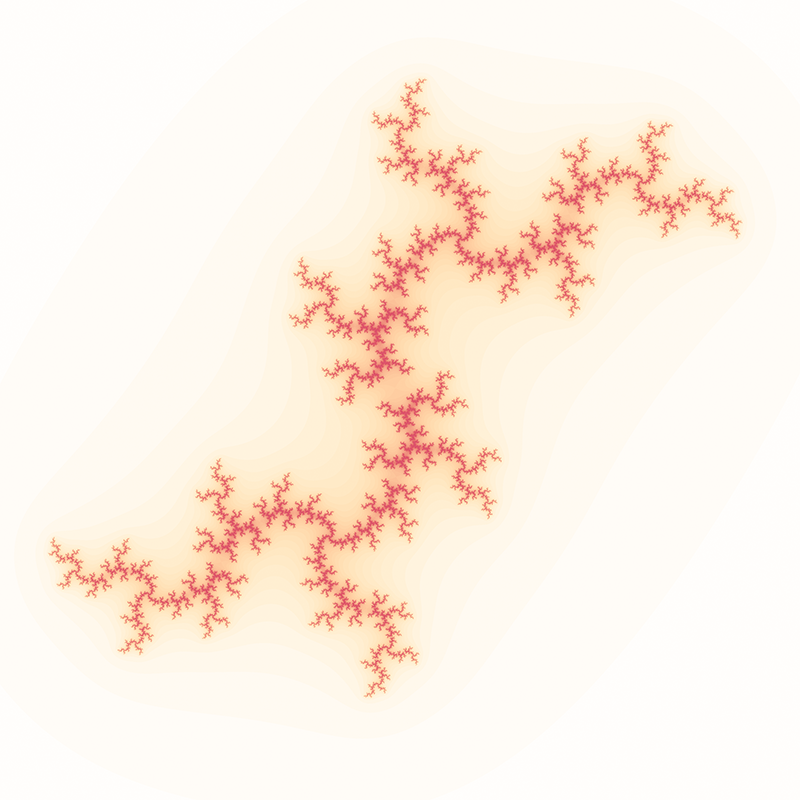} \\
	\includegraphics[width=0.23\linewidth]{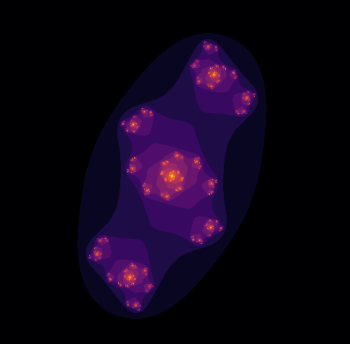}
	\qquad
	\includegraphics[width=0.23\linewidth]{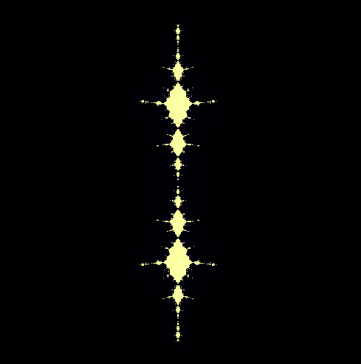}
	\caption{Some Julia sets of a family of quadratic polynomials,  $z^2+c_n$ (first row figures), and cubic polynomials, $z^3+c_n z$ (second row figures), for different ranges of parameters $c_n$.}\label{fig:julia}
\end{figure}

It is well-known in complex dynamical systems that, the Julia (denoted by $\mathcal{J}_{\omega}$) and Fatou (denoted by $\mathcal{F}_{\omega}$) sets are complementary sets in complex dynamics, defined by the behaviour of iterated complex functions. The Julia set represents points where the iterated function behaves chaotically, while the Fatou set contains points where the iterations are regular and predictable. They are fundamental to the study of fractal geometry and complex systems, and their properties have been explored extensively in mathematics. For more details on these aspects see for example \cite{Milnor, Beardon} on autonomous and \cite{Br, FS} for non-autonomous dynamical systems.  Figure \ref{fig:julia} illustrates an example of the Julia set of a random quadratic family of polynomials $z^2+c_n$. 
A key aspect explored in this context, is the stability of the Julia set associated with these maps. Our previous mathematical results \cite{stability} indicate that, by choosing the parameters $c_n$ randomly within a disk of radius $\delta < \delta_0\simeq 0.89$, the Julia set exhibits stable behaviour. 
This stability is crucial for the consistency of the generated pseudo-random key, and is particularly relevant in environments such as 5G networks, where small disturbances or noise can compromise the synchronization of the keys \cite{sun2020chaotic}. On the other hand, when $\delta > 3$, the system enters a highly chaotic regime, which is desirable for the generation of unique and ultra-secure keys \cite{alvarez2020some}.

We also present and discuss a stream cipher that combines fundamentals of random complex dynamical systems of the family of random cubic family of polynomials$f_{c_n}(z)=z^3+c_n z$, with modern cryptographic primitives (\texttt{HKDF} and \texttt{HMAC-SHA-256}). In fact, the algorithm generates a \emph{keystream} from orbits of an initial point $z_0$, with deterministic and reproducible sampling of the parameters $(c_n)$ through  a \texttt{DRBG HMAC}, \texttt{warm\_up} to reduce transients, and extraction by \texttt{HMAC} over binary representations of $f^k_{(c_n)}(z_0)$. Integrity protection is performed by \emph{Encrypt-the-MAC}. We present a design analysis, security considerations, and experimental results with \texttt{NIST SP 800-22} statistical tests. In a representative experiment, all reported tests passed.

\section{Preliminaries}
We devote this section to some preliminaries and basic results that we need during this article.
\subsection{\texttt{PRNG} and Chaotic Dynamical Systems}
In cryptography, a Pseudo-Random Number Generator (\texttt{PRNG}) is an algorithm designed to produce sequences of numbers that closely mimic true randomness, even though they are generated deterministically. While genuinely random numbers are impossible to achieve with an algorithm, cryptographic \texttt{PRNG}s are engineered to be highly unpredictable, resistant to pattern detection, and statistically indistinguishable from truly random sequences for all practical purposes. 

Key Characteristics of Cryptographic \texttt{PRNG}s are as follows:

(1) Determinism: At its core, a \texttt{PRNG} is a deterministic algorithm. It starts with a secret initial value known as a seed. Every subsequent number in the sequence is mathematically derived from the preceding one. This means that if you use the same seed, you'll always get the identical sequence of "random" numbers (see \cite{Stallings} for more details).

(2) Periodicity: All \texttt{PRNG}s eventually repeat their output sequence. However, a robust cryptographic \texttt{PRNG} boasts an extremely long period, making it computationally infeasible for an attacker to observe the repetition within any realistic timeframe (see \cite{Shallit}).

(3) Unpredictability: This is paramount for cryptographic security. Given a portion of the \texttt{PRNG}'s output, it should be computationally infeasible to predict any future numbers in the sequence (forward unpredictability) or any past numbers (backward unpredictability). This characteristic fundamentally differentiates cryptographic \texttt{PRNG}s from those used in non-security-critical applications like simulations (see \cite{Goldreich}).

(4) Statistical Randomness: The generated numbers must pass rigorous statistical tests for randomness. They should exhibit properties such as uniform distribution, a balanced ratio of zeros and ones, and an absence of discernible patterns that could be exploited by an adversary (see \cite{Nist1, Nist2, NIST:FIPS198-1:2008, NIST2010} for more details).

(5) Seed Secrecy and Entropy: The security of a cryptographic PRNG hinges on the secrecy and randomness of its initial seed. This seed must originate from a high-entropy source, meaning a source of genuine randomness, such as environmental noise, user input timings, or dedicated hardware random number generators to prevent attackers from guessing or deriving it (see \cite{Nist2}).

\subsubsection{Chaotic dynamical systems} 
In discrete time, even simple maps can be chaotic. For example, the logistic map $x_{n+1}=r x_n(1-x_n)$ or the Tent map are one-dimensional maps that rapidly mix points in $[0,1]$. The Tent map, defined by $T(x)=r\min(x,1-x)$, is widely used in cryptography for pseudo-random generation \cite{Al-Daraiseh}. 

Chaotic dynamical systems are characterized by their extreme sensitivity to initial conditions (the butterfly effect), resulting in unpredictable long-term behavior even in deterministic systems. This property makes them attractive candidates for building cryptographic \texttt{PRNG}s. When iterating over a chaotic mapping, a small difference in the initial state can lead to an exponential divergence in the generated sequences, creating a source of computational “randomness”.

However, classical maps also suffer drawbacks: limited parameter range and uneven distributions can limit security. For example, the authors in \cite{Al-Daraiseh} note that a $1$-dimensional Tent map has 
(i) only a restricted key range, (ii) periodic windows in its parameter space, and (iii) non-uniform output distributions.

\subsection{Random complex cubic polynomials}
The heart of our \texttt{PRNG} is the random iteration of the complex cubic polynomials $f_n(z) = z^3 + c_n z$, where the sequence of parameters $\omega=(c_n)$ are chosen randomly from a bounded Borel set in $\mathbb{C}$. 

The dynamics of these mappings are studied in the Riemann Sphere $\widehat{\mathbb{C}}=\mathbb{C}\cup\{\infty\}$, where $z$ is a complex number and $c_n$ is a complex parameter that varies with each iteration, as given in \eqref{eq:Fn}. 

The choice of parameters $c_n$ within a disk of radius $\delta$ controls the stability of the Julia set $\mathcal{J}_{\omega}$, note that the Julia set is the boundary of the filled Julia set $\mathcal{K}_{\omega}$ which is the set of points in $\mathbb{C}$ whose orbits remain bounded under the random iteration of $f_n(z)$. 

The following observation is a summary of results given in \cite{stability}.

\begin{remark}\cite{stability}\rm\label{rem:stable}
	Let $f_n(z)=z^3+c_n z$ with $(c_n)\in \overline{D}_{\delta}^{\mathbb{N}}$, where $\overline{D}_{\delta}$ is the closure of a disk of radius $\delta>0$ with center at the origin, i.e.
	\[
	\overline{D}_{\delta} = \{ c \in\mathbb{C}; \,\, |c|\leq \delta \}.
	\]
	We have:
	\begin{enumerate}
		\item \textbf{Stability ($\delta < \delta_0$):} For a radius $\delta$ smaller than a threshold $\delta_0 \simeq 0.89$, the Julia set is ``stable". This means that small variations in $(c_n)$ do not drastically change the structure of the Julia set, which translates into more consistent pseudo-random keys. This property is crucial for synchronization in noisy environments.
		\item \textbf{Instability ($\delta > 3$):} For $\delta > 3$, the system exhibits highly chaotic behavior. The orbits diverge rapidly, leading to the generation of more ``random" and unpredictable sequences, ideal for applications requiring high security and uniqueness of keys.
	\end{enumerate}
\end{remark}

It is worth noting that, the stability mentioned in Remark \ref{rem:stable} is particularly vital for technologies like 5G (the parameters $c_n$ act like noise of the system), where encryption keys must be synchronized between the sender and the receiver despite unavoidable noise or small disturbances. But, numerically, even if the starting point $z_0$ is chosen in the Julia set of $f_n$, it is difficult for its orbit to remain inside the Julia set under numerical iteration. This is because the Julia set is fractal, usually of measure zero and has no interior. Therefore, any small numerical ``error" will almost certainly push the orbit towards the Fatou set, where the dynamics are stable. Furthermore, in random dynamical systems, where the map $f_n$, changes at each step due to the variation of the parameters $c_n$, the Julia set itself may not be invariant in the classical sense. Despite this, the combination of $c_n$, a warm-up phase to allow the system to settle into complex behavior, and sensitivity to initial conditions in the map produce orbits with sufficient unpredictability for cryptographic applications.

\subsection{Hash-based Message Authentication Code}
A cryptographic hash function is a mathematical algorithm that takes an arbitrary block of data (the input, or ``message'') and outputs a fixed-size bit string, known as the \textit{hash value} \cite{Stallings:2017}. Unlike encryption, a hash function is a one-way function, i.e., it's computationally impossible to reverse the process, this means that, one can not reconstruct the original input data from its hash value.

For a hash function, $H$, to be considered cryptographically secure, it must satisfy several critical properties (the reader can find more details in  \cite{Menezes:1996, Schneier:1996} for instance):
\begin{enumerate}
	\item \textbf{Pre-image resistance (one-way property):} Given a hash value $h$, it is computationally impossible to find any input $M$ such that $H(M)=h$. This property prevents an attacker from recovering the original message from its hash.
	\item \textbf{Second pre-image resistance (weak collision resistance):} Given an input $M_1$, it is computationally impossible to find an input $M_2\neq M_1$ such that $H(M_1)=H(M_2)$. This property prevents an attacker from replacing an original message with a different one that produces the same hash.
	\item \textbf{Collision resistance (strong collision resistance):} It is computationally impossible to find \textit{any} two different inputs $M_1$ and $M_2$ such that $H(M_1)=H(M_2)$. This is the strongest property and implies second pre-image resistance.
\end{enumerate}
Here we represent some examples of existing Hash function methods.
\begin{itemize}
	\item[(i)] \textbf{\texttt{MD5} (Message-Digest Algorithm 5):} While historically popular, \texttt{MD5} is now considered cryptographically broken due to the discovery of practical collision attacks. It should not be used for security applications requiring collision resistance \cite{Wang:2005}.
	\item[(ii)] \textbf{\texttt{SHA-1} (Secure Hash Algorithm 1):} Similar to \texttt{MD5, SHA-1} has known collision vulnerabilities and is deprecated for most cryptographic uses by \texttt{NIST} and other standards bodies \cite{Nist1, Nist2, NIST:SP800-107r1:2012}.
	\item[(iii)] \textbf{\texttt{SHA-2} (Secure Hash Algorithm 2):} This family includes \texttt{SHA-256, SHA-384, SHA-512}, etc., differing in output hash length. \texttt{SHA-2} algorithms are widely used and remain cryptographically secure for now, although some long-term concerns exist about their reliance on the Merkle-Damgård construction.
\end{itemize}
A \textit{Hash-based Message Authentication Code} (\texttt{HMAC}) is a specific type of \textit{message authentication code} (\texttt{MAC}) that involves a cryptographic hash function (in our case, \texttt{SHA-256}), a secret key, an initial point $z_0$, iteration and warm-up numbers.

While a standard hash function only verifies that the data has not been altered, it does not provide authentication of the sender’s identity. If an attacker modifies a message and recalculates its hash, the recipient will still verify the authenticity of the altered message.

\texttt{HMAC} solves this problem by incorporating a secret key known only to the sender and the legitimate recipient. The \texttt{HMAC} process involves hashing the message twice with two different blocks derived from the ``secret key", making it dependent on both the ``message" and the shared ``secret keys". Only someone with the correct ``secret key" can calculate or verify the correct \texttt{HMAC} tag for a given message.

According to \cite{Bellare:1996}, \texttt{HMAC} ensures message authentication and data integrity. Also, it is resistance to length extension attacks which is a common vulnerability in simpler keyed hash schemes. A ``Deterministic Random Bit Generator" (\texttt{DRBG}), is an algorithm that produces a sequence of random bits based on an initial ``seed" value, and it can be reproduced if the same seed is used.

\subsection{Statistical Tests for Randomness}
The most significant area for improvement for cryptographic applications is to rigorously test the randomness of the generated key stream.
In this paper, first we evaluate some basic statistical tests, which are:
\begin{enumerate}
	\item \textbf{Chi-Square:} Evaluates the uniformity of the distribution of data. In cryptography, it is often applied to check whether the bits (or bytes) in a sequence occur with expected frequencies. In \cite{Knuth} it is shown that a $\chi^2$ value $<16.92$ indicates that the sequence is statistically uniform and there is no evidence of significant bias.
	
	\item \textbf{Autocorrelation:} The Autocorrelation test examines the dependence between elements of a sequence at different lags. In a truly random sequence, the current bits or bytes should have no correlation with the previous bits or bytes (\cite{Menezes:1996}).
	
	\item \textbf{Entropy:}  Entropy is a measure of the unpredictability or randomness of a source of information. In cryptography, it quantifies the uncertainty of each bit (or byte) generated.  For an ideally random sequence of bytes, where each byte from $0$ to $255$ has the same probability of occurring, the maximum entropy per byte is $8.0 = \log_2(256)$ bits. See \cite{NIST:SP800-107r1:2012, Shannon} for more details.
\end{enumerate}
To assess the quality of ``randomness" of a pseudorandom sequence, statistical tests based on \texttt{NIST SP 800-22} must be applied. These tests are:\\ 
\begin{enumerate}
	\item \textbf{Frequency (Monobit) Test:} Determines if the number of ones and zeros in the entire sequence is approximately equal.
	It calculates the number of ones and zeros and then uses a chi-squared or normal approximation to see if the counts are significantly different from what would be expected in a random sequence.
	\item \textbf{Frequency Within a Block Test:}
	Checks if the proportion of ones in non-overlapping blocks of the sequence is approximately 0.5. It's a localized version of the Monobit test.
	\item \textbf{Runs Test:}
	Determines if the number of ``runs" (consecutive sequences of identical bits) of various lengths is consistent with a random sequence. A run is an uninterrupted sequence of identical bits (e.g., 000 or 11).
	\item \textbf{Longest Run of Ones Test: }Determines if the length of the longest run of ones in the sequence is consistent with a random sequence.
	In fact, it identifies the longest consecutive sequence of '1's in the sequence and compares its length to the expected distribution for a random sequence of the same length.
	\item \textbf{Binary Matrix Rank Test:}
	Checks for linear dependencies within fixed-size matrices constructed from the sequence. It assesses the linear independence of fixed-length sub-strings within the sequence.
	\item \textbf{Discrete Fourier Transform (Spectral) Test:}
	Detects periodic features in the sequence that would indicate a deviation from randomness. It analyzes the distribution of peaks in the discrete Fourier transform of the sequence.
	\item \textbf{Non-Overlapping Template Matching Test:}
	Detects the presence of specific fixed-length patterns (templates) that occur too often or too rarely.
	It counts the occurrences of a set of pre-defined non-overlapping patterns (templates) in the sequence.
	\item \textbf{Overlapping Template Matching Test:}
	Similar to the non-overlapping test, but it allows templates to overlap. This can detect more subtle patterns.
	\item \textbf{Maurer's ``Universal Statistical" Test:}
	Detects patterns that are compressible, meaning they have a shorter description than a truly random sequence. It assesses the ``compressibility" of the sequence without relying on a specific compression algorithm.
	\item \textbf{Linear Complexity Test:}
	Determines the length of the shortest linear feedback shift register (LFSR) that can generate the sequence. A truly random sequence should have high linear complexity.
	\item \textbf{Serial Test:}
	Determines if the occurrences of all possible $m$-bit patterns (overlapping) of a specified length are approximately equally likely. It generalizes the frequency test to overlapping blocks of bits.
	
	\item \textbf{Approximate Entropy Test:}
	Assesses the ``predictability" of the sequence. Lower approximate entropy suggests more predictability and thus less randomness.
	\item\textbf{ Cumulative Sums Test:}
	Detects whether the number of ones and zeros in prefixes of the sequence deviates significantly from what would be expected. It checks for bias in partial sums of the sequence.
	\item \textbf{Random Excursions Test:}
	Detects deviations from the expected number of cycles in a random walk. It focuses on the number of times a random walk, created from the sequence, returns to the origin.
	\item \textbf{Random Excursions Variant Test:}
	Similar to the Random Excursions Test, but it focuses on the number of times the random walk touches various specific states (not just the origin).
	
\end{enumerate}

\section{Methodology: Implementation of the Chaotic Stream Cipher}
Information security is a fundamental pillar of digital communication, and stream ciphers play a crucial role in encrypting data in real time.
Chaotic systems, characterized by strong dependence on initial conditions and erratic behavior, are intrinsically suited for generating pseudorandom sequences. However, the isolated use of chaotic maps can be susceptible to attacks due to their deterministic nature and, sometimes, lack of a uniform bit distribution. 

In this section, we propose a hybrid model that addresses these limitations by combining the inherent unpredictability of a chaotic map with the provable security of standardized cryptographic primitives such as \texttt{HMAC} and \texttt{HKDF} to create a robust and secure keystream generator.

\begin{figure}[htbp]
	\centering
	\resizebox{\textwidth}{!}{
		\begin{tikzpicture}[>=latex, thick]
		
		\tikzset{
			block/.style={
				rectangle,
				rounded corners,
				minimum width=3cm,
				minimum height=1.2cm,
				draw=black,
				fill=blue!20,
				align=center
			}
		}
		
		\node[block] (key) {Secret Key + IV};
		
		\node[block, right=2cm of key] (hkdf) {\texttt{HKDF}};
		
		\node[block, above right=1cm and 1.5cm of hkdf] (hmac) {\texttt{HMAC-DRBG}};
		\node[block, below right=1cm and 1.5cm of hkdf] (chaos) {Chaotic map $z^3 + c_n z$};
		
		\node[block, right=6.5cm of hkdf] (prng) {\texttt{PRNG}\\(Keystream)};
		
		\node[block, below=3cm of hkdf] (msg) {Message};
		\node[block, right=3cm of msg] (xor) {XOR};
		\node[block, right=1.5cm of xor] (cipher) {Cipher};
		
		\draw[->] (key) -- (hkdf);
		
		\draw[->] (hkdf) -- (hmac);
		\draw[->] (hkdf) -- (chaos);
		
		\draw[->] (hmac) -- (prng);
		\draw[->] (chaos) -- (prng);
		
		\draw[->] (msg) -- (xor);
		\draw[->] (prng) -- (xor);
		\draw[->] (xor) -- (cipher);
		
		\end{tikzpicture}
	}
	\caption{Block diagram of a key and cipher generator.}
	\label{fig:diagram}
\end{figure}
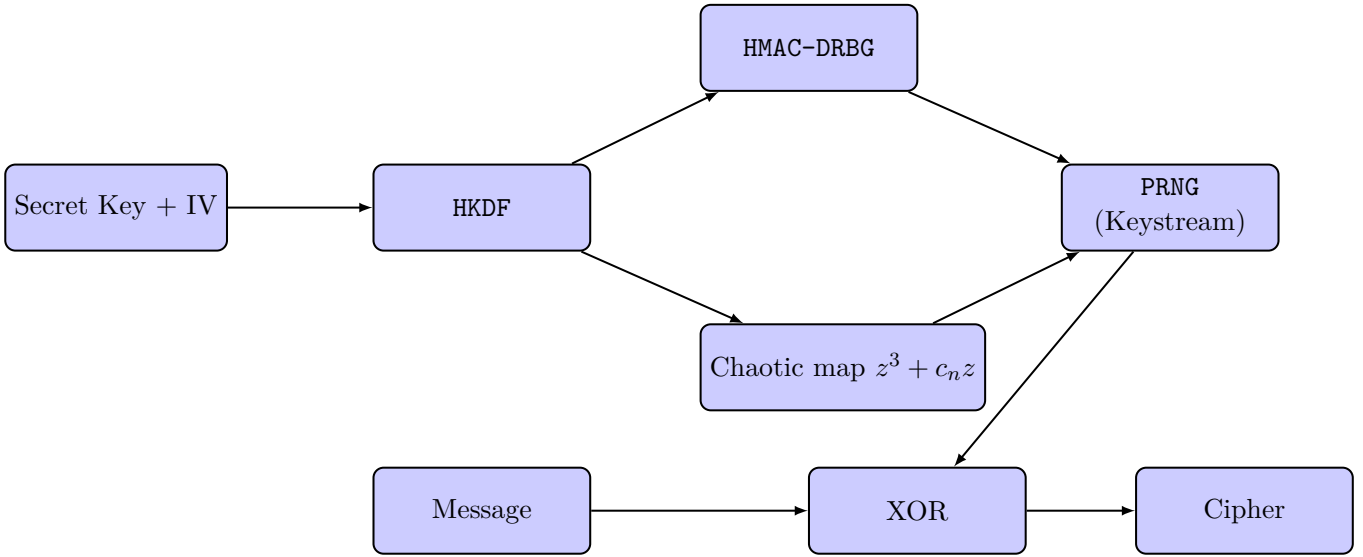

The system is built on three main components: the \textit{Cryptographic Randomness Generation}, the \textit{Chaotic Stream Cipher} and the \textit{Encryption} Scheme. Figure \ref{fig:diagram} illustrates a block diagram of this system.
\bigskip

\subsection{Cryptographic Random Generation}
To ensure system integrity, all randomness sources are cryptographically secure. The \texttt{secure\_bytes()} function uses the operating system's \texttt{os.urandom} interface to obtain high-quality bytes, ensuring that \texttt{IV}s (unique numbers per use) and keys are generated from an unpredictable source. The \texttt{secrets} module is used to avoid modulo bias in sampling operations, a critical security requirement.
\subsection{Chaotic Keystream Generator via random cubic polynomials}
The heart of the system is the keystream generator, based on a random cubic chaotic map:
\[f_{c_n}(z)=z^3+c_n z,\]
where $z$ is an initial complex number and the sequence of random complex parameters $(c_n)$ is chosen from a disc of radius $\delta$. The security of this generator is guaranteed by a secure seeding process:
\begin{enumerate}
	\item \textbf{Key Derivation:} A master key and an IV are passed to \texttt{HKDF-SHA256}, which derives two cryptographically independent subkeys: one for keystream generation and one for authentication (secret key). The sender and recipient of the message must share a pre-established secret key. It is crucial that this key remains secure. The recipient can derive the initialization vector (IV) from the encrypted message itself, eliminating the need for it to be transmitted separately.   
	\item \textbf{Seed:} An \texttt{HMAC-DRBG} is used to deterministically generate the initial map parameters, $c_n$ and $z_0$.
	This method ensures that the same set of key and IV always produces the same keystream, which is essential for decryption.
	\item \textbf{Warm Up:} The chaotic map is iterated \texttt{WARM\_UP} times without producing any output. This ``warm up" eliminates the influence of the initial conditions and forces the system to converge to a truly chaotic state.
	\item \textbf{Keystream Production:} Every three iterations, the system state \texttt{(z.real, z.imag)} and a counter are concatenated and processed by an \texttt{HMAC} to produce a 32-byte keystream block. This step is crucial for obscuring the details of the map's floating-point state and ensuring a uniform, high-entropy bit distribution in the output stream.
\end{enumerate}

\subsection{Encryption and Decryption}
The encryption scheme uses \texttt{encrypt} and \texttt{decrypt} functions, which are a security standard to ensure both confidentiality and data integrity. 

The \texttt{encrypt} function takes the following variables as input: \texttt{plaintext}, \texttt{key}, \texttt{ad} (associated data), and an optional \texttt{IV} (initialization vector). It returns a single bytes object, which is a concatenation of the \texttt{IV, ciphertext}, and \texttt{tag}.

Please note the following requirements for the variables:\\
\texttt{key}: A 32+ byte key is recommended and will be split using \texttt{HKDF}.\\
\texttt{ad}: This associated data is not encrypted but is authenticated.\\
\texttt{IV}: If not provided, a random 16-byte IV will be generated. The \texttt{IV} must be unique for each key.\\
It is worth noting that, the plaintext is combined with the keystream via \texttt{XOR} operation to produce the \texttt{ciphertext}.

\begin{verbatim}
def encrypt(plaintext: bytes,
key: bytes,
ad: bytes = b"",
iv: Optional[bytes] = None) -> bytes:

if iv is None:
iv = os.urandom(16)

# Derive independent subkeys for stream & MAC from (key, iv, ad)
kdf_out = hkdf_sha256(ikm=key, salt=iv, info=b"split"+ad, length=64)
stream_key, mac_key = kdf_out[:32], kdf_out[32:]

keystream = _keystream_from_map(stream_key, iv, ad, len(plaintext))
ciphertext = bytes(p ^ k for p, k in zip(plaintext, keystream))

mac = hmac.new(mac_key, ad + iv + ciphertext, hashlib.sha256).digest()

# Concatenate nonce, ciphertext, and tag for transmission
return iv + ciphertext + mac
\end{verbatim}

The \texttt{decrypt} function receives the concatenated message (i.e., \texttt{encrypted\_message: bytes}), a \texttt{key}, and optional \texttt{ad} (associated data), and returns the decrypted \texttt{plaintext}. To perform this, the function first extracts the \texttt{IV}, \texttt{ciphertext}, and \texttt{tag} from the concatenated message. It then uses these values to verify the \texttt{HMAC}. Only after a successful authentication is the \texttt{ciphertext} decrypted to recover the original message.

\begin{verbatim}
def decrypt(encrypted_message: bytes,
key: bytes,
ad: bytes = b"") -> bytes:

# Extract IV, ciphertext, and tag from the concatenated message
# Assuming IV is 16 bytes and tag is 32 bytes (SHA-256 output size)
iv_size = 16
tag_size = 32

if len(encrypted_message) < iv_size + tag_size:
raise ValueError("Encrypted message is too short.")

iv = encrypted_message[:iv_size]
ciphertext = encrypted_message[iv_size:-tag_size]
tag = encrypted_message[-tag_size:]


kdf_out = hkdf_sha256(ikm=key, salt=iv, info=b"split"+ad, length=64)
stream_key, mac_key = kdf_out[:32], kdf_out[32:]

mac_check = hmac.new(mac_key, ad + iv + ciphertext, hashlib.sha256)
.digest()
if not hmac.compare_digest(tag, mac_check):
raise ValueError("Authentication failed")

keystream = _keystream_from_map(stream_key, iv, ad, len(ciphertext))
plaintext = bytes(c ^ k for c, k in zip(ciphertext, keystream))
return plaintext
\end{verbatim}

\section{Statistical evaluation \texttt{(NIST SP 800-22)}}
To evaluate the quality of the randomness of the key stream generated by the random cubic map-based \texttt{PRNG}, standard statistical tests were applied. We report in Table \ref{tab:nist}, $p$-values given for a sequence of $n=6480$ bits. A test is ``passed” when the $p$-value exceeds the chosen threshold (typically $\alpha=0.01$).

\begin{table}[h]
	\begin{center}
	\begin{tabular}{|l|l|c|}
		\hline
		Test & p-value & Pass \\
		\hline
		\hline
		Frequency (Monobit) & 0.6629 & Yes \\
		Block Frequency & 0.6856 & Yes \\
		Cumulative Sums (fwd) & 0.4257 & Yes \\
		Cumulative Sums (bwd) & 0.8063 & Yes \\
		FFT & 0.8192 & Yes \\
		Approximate Entropy & 0.0184 & Yes \\
		Linear Complexity & 0.4102 & Yes \\
		Longest Run of Ones & 0.4121 & Yes \\
		Non-overlapping Templates & 0.8556 & Yes \\
		Overlapping Templates & 0.3583 & Yes \\
		Random Excursions & 0.4158 & Yes \\
		Random Excursions Variant & 0.4226 & Yes \\
		Rank ($32\times 32$) & 0.2248 & Yes \\
		Runs & 0.7222 & Yes \\
		Serial (m=3) & 0.4462/0.1731 & Yes/Yes \\
		Universal (Maurer) & 0.0931 & Yes \\
		\hline
	\end{tabular}
	\vspace{0.2cm}
	\caption{Summary results of \texttt{NIST SP 800-22} tests (sequence with $n=6480$ bits).}
	\label{tab:nist}
	\end{center}
\end{table}

\subsection{Some graphical tools}
\begin{figure}[h!]
	\centering
	\begin{minipage}{0.45\linewidth}
		\centering
		\includegraphics[width=\linewidth]{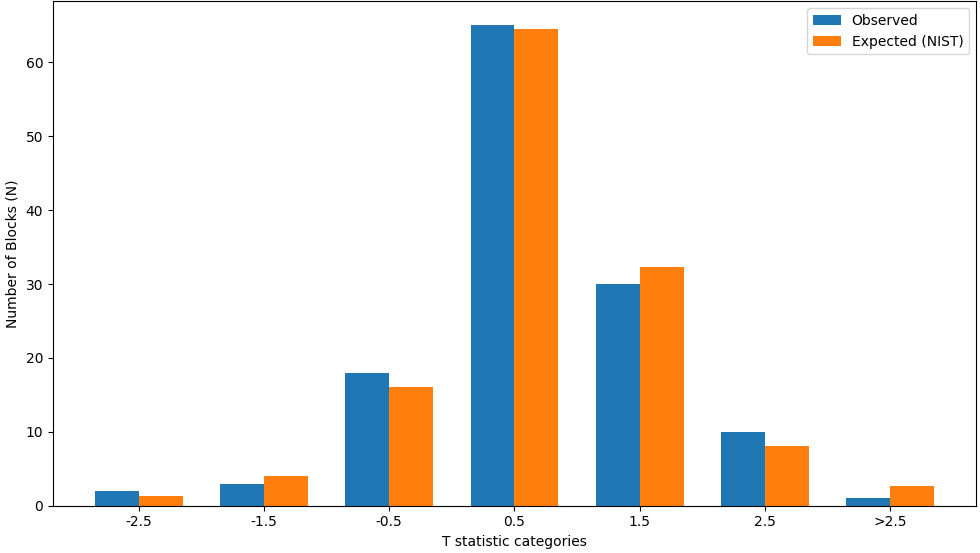}
		\caption*{Linear Complexity Test.}
	\end{minipage}
	\hfill
	\vspace{0.5cm}
	\begin{minipage}{0.45\linewidth}
		\centering
		\includegraphics[width=\linewidth]{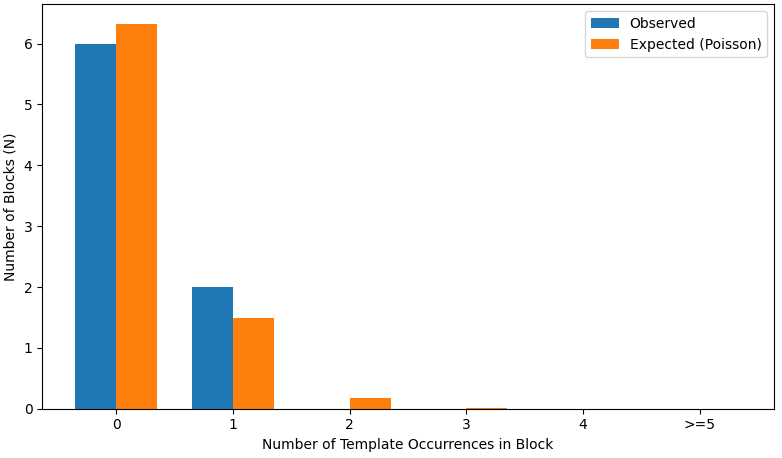}
		\caption*{Overlapping Template Test.}
	\end{minipage}

	\begin{minipage}{0.5\linewidth}
		\centering
		\includegraphics[width=\linewidth]{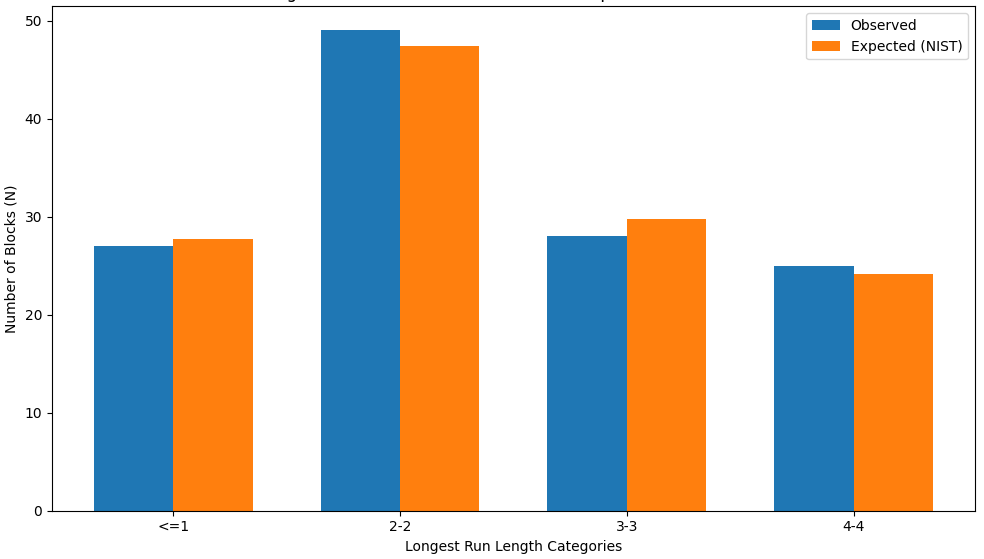}
		\caption*{Longest Run Test.}
	\end{minipage}
	\hfill
	\vspace{0.5cm}
	\begin{minipage}{0.4\linewidth}
		\centering
		\includegraphics[width=\linewidth]{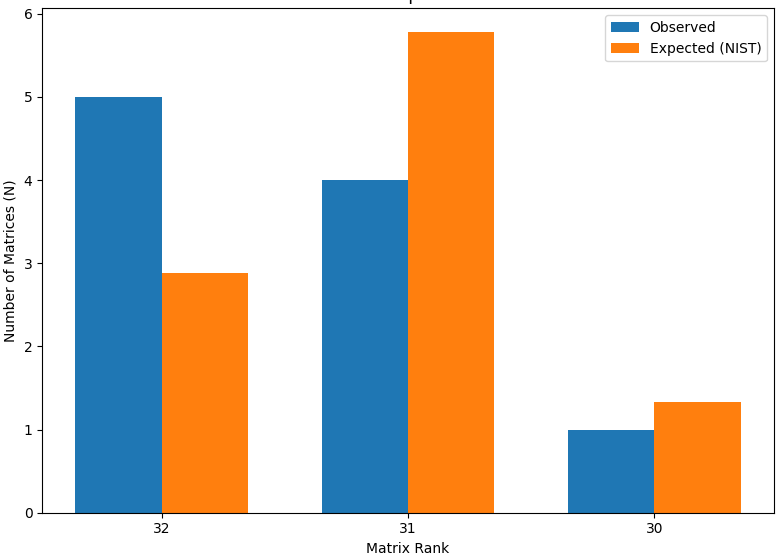}
		\caption*{Rank Test.}
	\end{minipage}
	\caption{Comparision of observed results (blue histograms) with expected \texttt{NIST} tests (orange histograms).}\label{fig:CO} 
\end{figure}

\begin{figure}[hbt!]
	\begin{minipage}{0.45\linewidth}
		\centering
		\includegraphics[width=\linewidth]{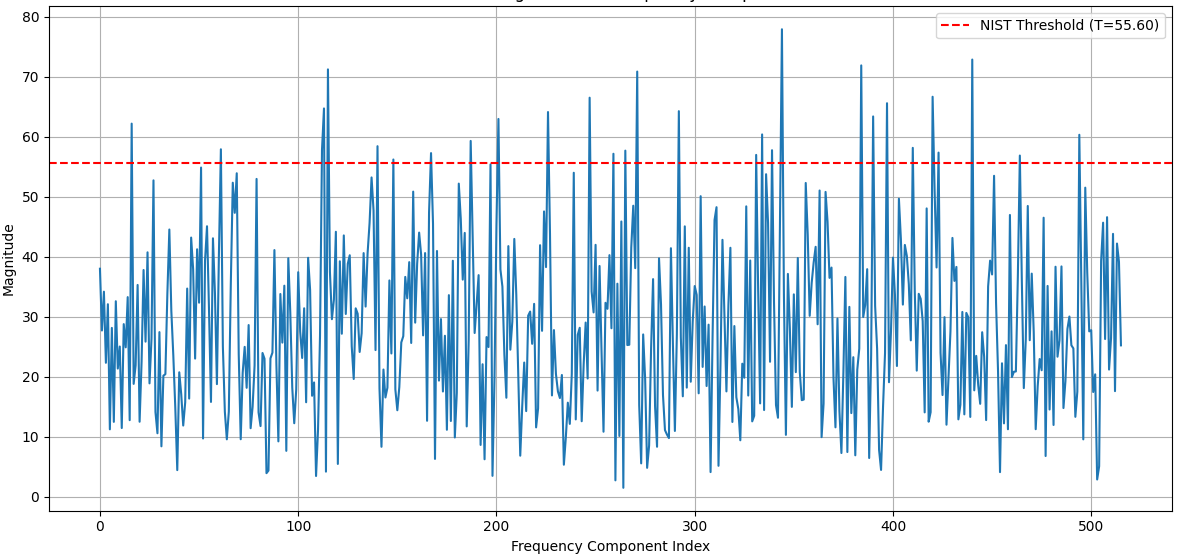}
		\caption*{FFT Test.}
	\end{minipage}
	\hfill
	\vspace{0.5cm}
	\begin{minipage}{0.45\linewidth}
		\centering
		\includegraphics[width=\linewidth]{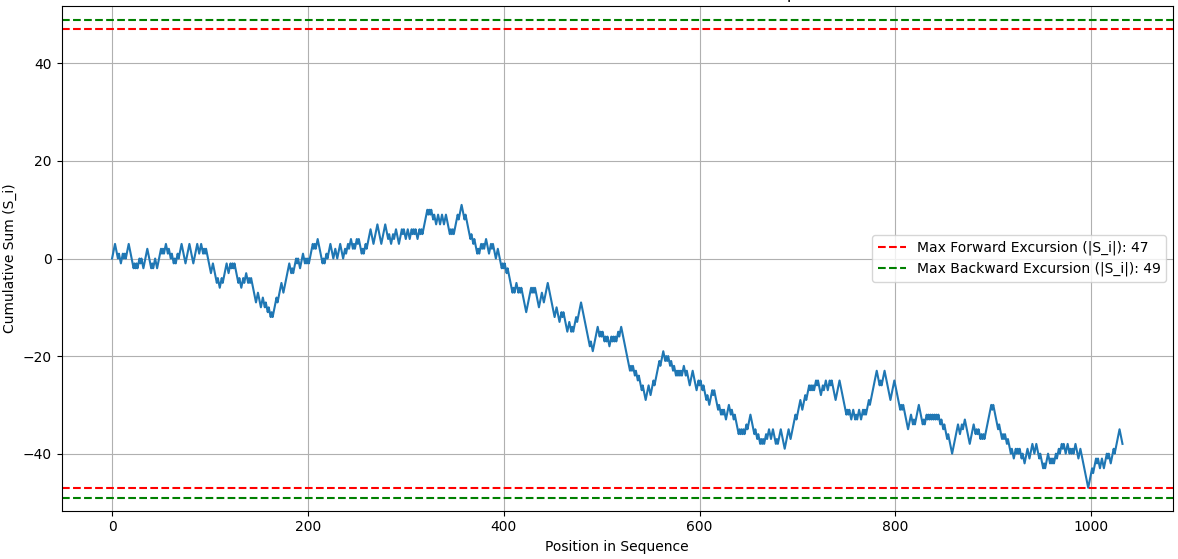}
		\caption*{Cumulative Sums Test.}
	\end{minipage}
	\caption{}
	\label{fig:combined_fft}
\end{figure}

\begin{figure}[hbt!]  
\begin{minipage}{0.36\linewidth}
	\centering
	\includegraphics[width=\linewidth]{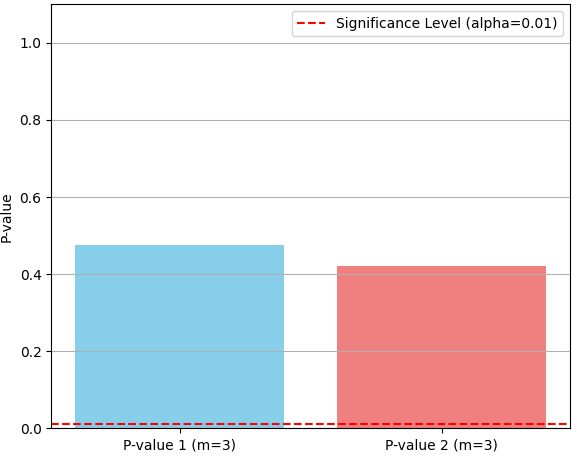}
	\caption*{Serial Test.}
\end{minipage}
\hspace{2cm}
\vspace{0.5cm}
\begin{minipage}{0.45\linewidth}
	\centering
	\includegraphics[width=\linewidth]{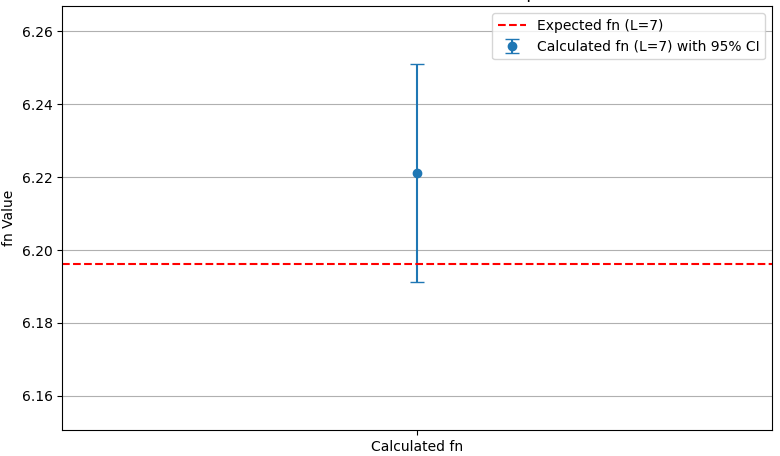}
	\caption*{Universal Test.}
\end{minipage}
\caption{}
\label{fig:combined_fft2}
\hfill
\vspace{0.5cm}
\end{figure}

To have a better visualization of statistical tests some graphical tools has been implemented. While they are not rigorous statistical tests in themselves, they provide valuable intuition about the randomness of a sequence (see Figures \ref{fig:CO},\ref{fig:combined_fft}, \ref{fig:combined_fft1} and \ref{fig:combined_fft2}).

\section{\texttt{ENT} Statistical Test}
\texttt{ENT} (also known as the Fourmilab Random Sequence Tester) is a statistical test suite that evaluates the quality of random or pseudorandom number sequences by checking for deviations from perfect randomness. It uses several statistical measures, including entropy, chi-square, arithmetic mean, Monte Carlo Pi estimation, and serial correlation, to determine if the sequence is sufficiently unpredictable for applications like cryptography, simulations, and statistical sampling. Table \ref{tab:ent} illustrates the results of these tests.

\begin{table}[h]
	\begin{center}
	\begin{tabular}{|l|l|c|}
		\hline
		Statistical Test & Conditions & Results \\
		\hline
		\hline
		Entropy &  $\approx 8$ (for one byte) & 6.6894\\
		\hline
		Optimum Compression &  $\approx 1$ & $0.8362$ \\
		\hline
		Chi-Square & $10\%-90\%$ & $93.15\%$ \\
		\hline
		Arithmetic Mean & 127.5 & 119.9302 \\
		\hline
		Monte Carlo Value for $\pi$ &  0\% error & 1.4613\% \\
		\hline
		Serial Correlation Coefficient &  0  & -0.0876 \\
		\hline
	\end{tabular}
	\caption{Summary results of \texttt{ENT} statistical tests.}
	\label{tab:ent}
	\end{center}
\end{table}

There are some observations about the results of Table \ref{tab:ent} that we should consider:
\begin{enumerate}
	\item \textbf{Entropy:} This measures the randomness of the data in bits per byte. A value close to 8.0 bits per byte indicates high randomness, as each byte is nearly unpredictable. our result of $6.69$ is reasonably high, suggesting good randomness at the byte level.
	\item \textbf{Optimum Compression Ratio:} This is related to entropy and represents the theoretical limit of compression for the data. A value close to $1.0$ means the data is not very compressible, which is expected for random data. Our result of $0.84$ supports this.
	\item \textbf{Chi-squared statistic:} This tests the uniformity of the byte distribution. For truly random data, the expected count for each byte value ($0-255$) is the total number of bytes divided by $256$. A high $p$-value (like in our case $0.9315$) means that the observed distribution of byte values in the encrypted message is very close to what we would expect from a perfectly uniform random sequence.
	\item \textbf{Arithmetic Mean:} For truly random bytes $(0-255)$, the expected arithmetic mean is $(0 + 255) / 2 = 127.5$. Our result of $119.93$ is close to this expected value, which is a good sign.
	\item \textbf{Monte Carlo $\pi$:} This test uses pairs of bytes as coordinates to approximate the value of $\pi$. For random data, the approximation should be close to the actual value of $\pi$. Our result $1.4613\%$ of error. Note that, the accuracy of this test depends on the sample size, and our data length resulted in a smaller sample size, which leads to bigger error.
	
	\begin{figure}[hbt!]  
		\begin{minipage}{0.39\linewidth}
			\centering
			\includegraphics[width=\linewidth]{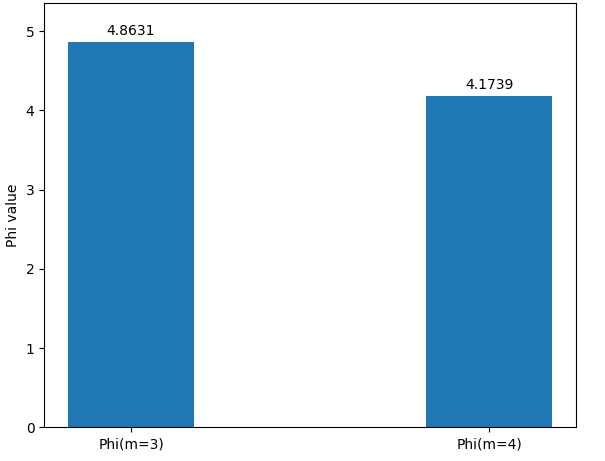}
			\caption*{Approximate Entropy Test.}
		\end{minipage}
		\hfill
		\vspace{0.5cm}
		\begin{minipage}{0.385\linewidth}
			\centering
			\includegraphics[width=\linewidth]{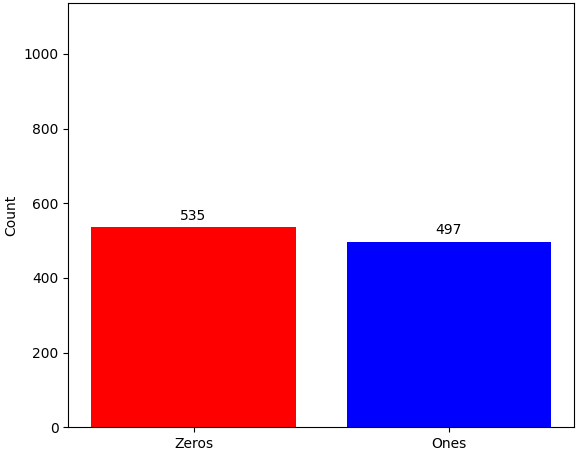}
			\caption*{Frequency Monobit Test.}
		\end{minipage}
		\hfill
		\vspace{0.5cm}
		\begin{minipage}{0.4\linewidth}
			\centering
			\includegraphics[width=\linewidth]{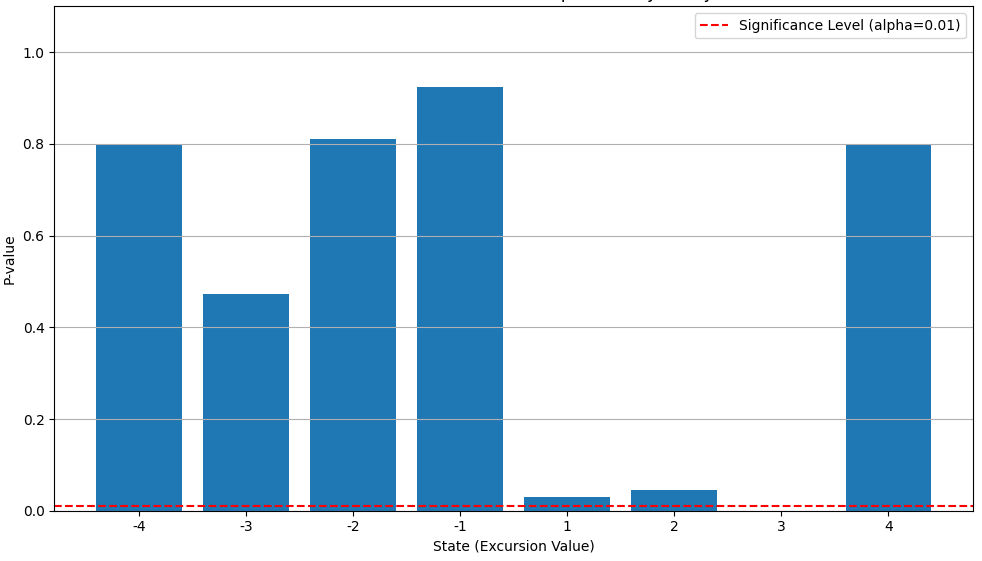}
			\caption*{Random Excursions.}
		\end{minipage}
		\hfill 
		\vspace{0.5cm}
		\begin{minipage}{0.49\linewidth}
			\centering
			\includegraphics[width=\linewidth]{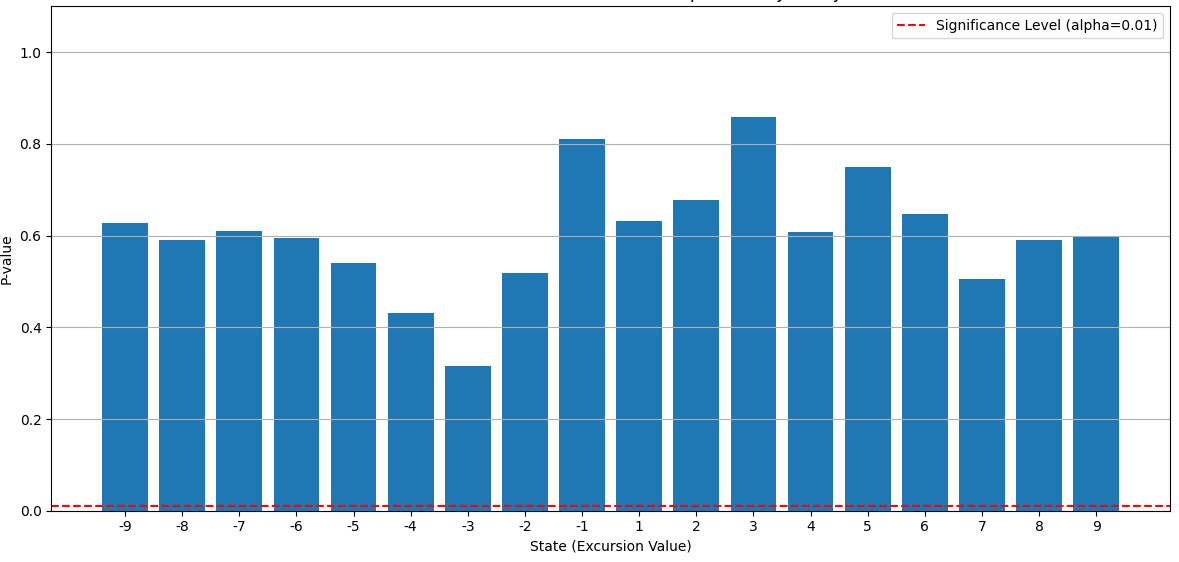}
			\caption*{Random Excursions.}
		\end{minipage}
		\caption{}
		\label{fig:combined_fft1}
		\hfill
		\vspace{0.5cm}
	\end{figure}
	\item \textbf{Serial Correlation Coefficient:} This measures the correlation between adjacent bytes. For truly random data, the serial correlation should be close to $0$. Our result indicates the degree of linear relationship between consecutive bytes. A value further from 0 suggests some level of predictability between adjacent bytes.
\end{enumerate}

\begin{remark}
	Based on the \texttt{ENT} results, the encrypted message appears to have some good statistical properties of randomness, particularly in terms of entropy, arithmetic mean and the Chi-squared statistic. The serial correlation coefficient might warrants further investigation to achieve a very high standard of randomness.
\end{remark}

\section{The \texttt{U01} test suite and Some Discussion}
According to the statistical analysis given in the previous sections, our \texttt{PRNG} model passes statistical tests and is well balanced. But cryptographic security requires deeper analysis such as, \texttt{U01} test suite and
other additional security properties like what is done for recent \texttt{PRNG}s for example, \texttt{ChaCha20}, \texttt{SHAKE-128} and \texttt{AES-CTR DRBG} (see \cite{chacha20, SHA3, Barker}).

The \texttt{U01} test suite refers to a set of statistical tests for randomness developed by \texttt{NIST} and designed to determine the random behavior of \texttt{PRNG} (see \cite{u01}). 
These tests are designed to evaluate sequences of bits produced by random number generators. The suite includes tests like the \texttt{NIST} tests (Frequency, Block Frequency, Cumulative Sums, Runs, Longest Run, Rank, FFT, Non-overlapping Template, Overlapping Template, Universal, Approximate Entropy, Serial, Random Excursions, and Random Excursions Variant) that we observed before. The \texttt{NIST} tests passed for smaller bit sequences (like the 1032-bit sequence) but for \texttt{U01} tests we need running the full \texttt{U01} test suite on sequences of big sizes (like $2^{20}$ or $2^{30}$-bits) and it failed at Approximate Entropy, Longest Run, and Overlapping Template tests for the larger $2^{20}$-bit sequence. 

The failures in tests like Approximate Entropy, Longest Run, and Overlapping Template for the $2^{20}$-bit sequence indicate that while the chaotic map's output might look random and pass basic checks on smaller scales, it contains subtle statistical non-uniformities that become apparent at larger scales. These biases might be inherent to the specific chaotic map or the way its output is processed into a bit stream.
Here's a breakdown of what we attemp to improve mixing:

We modified \texttt{\_keystream\_from\_map} to accumulate the packed real/imaginary parts of $z$ for a few iterations ($5$, then $10$) before hashing. These modifications did not resolve the failures in the Approximate Entropy Test, Longest Run Test, and Overlapping Template Test.
This indicated that simply concatenating and hashing the raw chaotic output for a few steps wasn't effectively removing all the statistical biases detected by the failing tests.

We also modified \texttt{\_keystream\_from\_map} to maintain a running \texttt{SHA-256} hash, updating it with the packed real/imaginary parts of $z$ after each chaotic map iteration, and then hashing the digest of this running hash to produce keystream blocks. This approach also did not resolve the failures in the Approximate Entropy Test.
The consistent failures in this test indicates that further work would be needed to make this specific chaotic map-based generator pass all the \texttt{NIST} statistical randomness tests.


\section{Conclusion and Future work}
This work demonstrated the practicability of a symmetric stream cipher based on random cubic polynomials in the complex plane, combined with \texttt{HMAC-SHA256} authentication. The approach of dynamically varying the parameters $c_n$ proved to be effective in inducing and controlling the chaotic behavior of the system, generating pseudo-random key streams with good statistical properties (for low bit sequences).


%
%
%

\subsection*{Acknowledgements}
The authors would like to thank Dr. Rodrigo Abarzúa (Universidad de Santiago - Chile) and Dr. Farid Tari (Universidade de São Paulo - Brazil) for their valuable comments and suggestions. 

The first author was partially supported by FAPEMIG APQ-02375-21 and BPD-00761-22 and RED-00133-21.	The second author was supported by Mathamsud Project TOMCAT 22-MATH-10 and ANID-FONDECYT 1230807. The forth author is supported by FAPEMIG post-doctoral scholarship with process number BPD-00761-22.

\end{document}